\documentclass[sigconf]{acmart}

\usepackage{subfigure}
\usepackage{graphicx}
\usepackage{epsfig}
\usepackage{color}
\usepackage[ruled,vlined]{algorithm2e}
\usepackage{amsmath,amssymb,amsfonts}

\begin{document}

\copyrightyear{2019}
\acmYear{2019}
\setcopyright{acmlicensed}
\acmConference{Conference}{}{Date, Year, Location.}
\acmPrice{15.00}
\acmDOI{http://dx.doi.org/DOI}
\acmISBN{ISBN ISBN}

\fancyhead{}
%\settopmatter{printacmref=false, printfolios=false}

%IF NEED BE.
% --- End of Author Metadata ---

\title{Computational Thinking with the Web Crowd using Code{\em Mapper}}
\titlenote{Research was supported in part by National Science Foundation Grant DRL 1515550.}
%\subtitle{Extended Abstract}
%\subtitlenote{The full version of the author's guide is available as
%  \texttt{acmart.pdf} document}

\author{Patrick Vanvorce}
%\orcid{0000-0002-3124-3780}
\affiliation{%
  \institution{Department of Computer Science}
  \streetaddress{University of Idaho, USA}
%  \city{Moscow}
%  \state{Idaho}
%  \postcode{83844, USA}
}
\email{(vanv1995@vandals.uidaho.edu}

\author{Hasan M. Jamil}
%\orcid{0000-0002-3124-3780}
\affiliation{%
  \institution{Department of Computer Science}
  \streetaddress{University of Idaho, USA}
%  \city{Moscow}
%  \state{Idaho}
%  \postcode{83844, USA}
}
\email{jamil@uidaho.edu}

\renewcommand{\shortauthors}{Hasan Jamil}
\renewcommand{\shorttitle}{Computational Thinking with the Web Crowd using Code{\em Mapper}}

\begin{abstract}
It has been argued that {\em computational thinking} should precede computer programming in the course of a career in computing. This argument is the basis for the slogan ``logic first, syntax later" and the development of many cryptic syntax removed programming languages such as Scratch!, Blockly and Visual Logic. The goal is to focus on the structuring of the semantic relationships among the logical building blocks to yield solutions to computational problems. While this approach is helping novice programmers and early learners, the gap between computational thinking and professional programming using high level languages such as C++, Python and Java is quite wide. It is wide enough for about one third students in first college computer science classes to drop out or fail. In this paper, we introduce a new programming platform, called the Code{\em Mapper}, in which learners are able to build computational logic in independent modules and aggregate them to create complex modules. Code{\em Mapper} is an abstract development environment in which rapid visual prototyping of small to substantially large systems is possible by combining already developed independent modules in logical steps. The challenge we address involves supporting a visual development environment in which ``annotated code snippets" authored by the masses in social computing sites such as SourceForge, StackOverflow or GitHub can also be used as is into prototypes and mapped to real executable programs. Code{\em Mapper} thus facilitates soft transition from visual programming to syntax driven programming without having to practice syntax too heavily.
\end{abstract}

\begin{CCSXML}
<ccs2012>
<concept_id>10003752.10010124</concept_id>
<concept_desc>Theory of computation~Semantics and reasoning</concept_desc>
<concept_significance>500</concept_significance>
</concept>
<concept>
<concept_id>10003752.10003809</concept_id>
<concept_desc>Theory of computation~Design and analysis of algorithms</concept_desc>
<concept_significance>300</concept_significance>
</concept>
<concept>
<concept_id>10003120.10003130.10003131.10003234</concept_id>
<concept_desc>Human-centered computing~Social content sharing</concept_desc>
<concept_significance>100</concept_significance>
</concept>
<concept>
<concept_id>10003120.10003130.10003131.10003235</concept_id>
<concept_desc>Human-centered computing~Collaborative content creation</concept_desc>
<concept_significance>100</concept_significance>
</concept>
<concept>
<concept_id>10003120.10003130.10003233.10003288</concept_id>
<concept_desc>Human-centered computing~Blogs</concept_desc>
<concept_significance>100</concept_significance>
</concept>
</ccs2012>
\end{CCSXML}

\ccsdesc[500]{Theory of computation~Semantics and reasoning}
\ccsdesc[300]{Theory of computation~Design and analysis of algorithms}
\ccsdesc[100]{Human-centered computing~Social content sharing}
\ccsdesc[100]{Human-centered computing~Collaborative content creation}
\ccsdesc[100]{Human-centered computing~Blogs}

\keywords{Computational thinking;
imperative programming;
rapid prototyping;
crowd-computing;
conceptual programming}

\maketitle

\section{Introduction}
The differences between computing and computational thinking are significant and can be explained in a number of ways. According to Jeannette Wing \cite{WingJ2016}, "computational thinking confronts the riddle of machine intelligence: What can humans do better than computers, and what can computers do better than humans? Most fundamentally, it addresses the question: What is computable? Today, we know only parts of the answers to such questions." In particular, computational thinking is 1) conceptualizing, not programming, 2) a way that humans, not computers, think, 3) complements and combines mathematical and engineering thinking, and 4) it is ideas, not artifacts. Our Mind{\em Reader}  project \cite{MindReader-ICALT-2017} aims at implementing these ideas into a system for online autonomous learning.

Millions of K-12 students by now are conversant in the graphical language Scratch (developed at MIT). However, the language is incredibly clunky if one wants to go beyond what it is designed to do (make simple games). A derivative language, Snap!  (developed at UC Berkeley), expands its horizons, making functions practical, enabling object oriented programming, and fixing other deficiencies in Scratch. Yet, many educators believe that it is a fantasy to expect the masses to transition from Scratch (or Snap!) to what we now think of as conventional programming languages such as C++, Python or Java. A more plausible future is for specialized languages to spring up that make use of the conventions of Scratch/Snap! but incorporate the knowledge and concepts of a field of interest to a target audience. It isn't difficult to see a world where almost everyone speaks Scratch (like almost everyone knows algebra), and when we go into, for example, molecular biology, we naturally adopt Scratch/Genome, or if one wants to go into accounting, she adopts Scratch/Accounting. In other words, a discipline specific toolbox or plug-in will tailor how the programming environment would interface with its users.

While comparative student-learning behaviors in different STEM (Science, Technology, Engineering and Mathematics) areas is unknown, anecdotally it is believed that biologists usually prefer abstract graphical tools more than physicists or mathematicians who prefer details. So, it is not surprising that computational physicists insist on a more hands on computer programming experience than computational biologists advocating visual or NLP based programming. We believe that students in different disciplines from different background have unique learning trait and learn differently. Thus, contrary to the idea that one language with discipline specific toolboxes or plugins is the way forward, we assemble a constellation of programming environments in Mind{\em Reader} that is discipline agnostic, but learning style specific, that will allow a learner to move between the environments based on her comfort zone. We believe by keeping in view what computational thinking is about, and learners' individuality, we are able to design a truly impactful system to support the vision of the educational standards such as ``Common Core State Standards Initiative" (CCSS) \cite{CCSS} and ``Next Generation Science Standard" (NGSS) \cite{NGSS}.

\section{Social Programming using Crowd-Sourced Modules}

Programming can be challenging for young programmers and STEM learners due to a host of reasons with abstract thinking, complex syntax, mapping seemingly simple steps into algorithms, etc. being the leading ones. For K-12 and early college STEM learners, a substantial number of online teaching, tutoring and assessment systems are emerging with various degrees of impact. In most part, these systems and tools have done little to reduce high drop-out rates \cite{WatsonL14}, or to aid learning, and resource strapped institutions continue to struggle to find better ways to train new century workforce conversant in computing \cite{VihavainenAW14}.

One of the critical decisions educators have been grappling with is how much exposure to programming is enough to make students interested and ultimately proficient in professional system building \cite{PorterGMS13}, and what languages to choose for this purpose \cite{KunkleA16}. While computational thinking is not specific about any language, it is imperative that one or a set of programming platforms, e.g., block based (e.g., Scratch, Snap!, Alice), declarative (e.g., Prolog, Clasp), imperative (e.g., C++, Java, Python), visual (e.g., Visual Logic, Cameleon, Snap!), programming by example (e.g., Foofah, BlinkFill) or natural language (e.g., AppleScript, {\sc Metafor}, Flip) among many other possible alternatives, need to be chosen. While the answers to these questions are still outstanding, most programs use an imperative language as the main platform even if they start off with a language such as Scratch.

It is worth noting that most of the programming languages capture the basic constructs of imperative programming -- assignment and computation, decision, and iteration. The block and frame based languages primarily adopt a {\em logic first, syntax later} approach \cite{KollingBA15}. Block languages aim to emphasize conceptual clarity, ease of coding and simplicity and thereby deemphasize programming rigor, computational power and other essential features imperative languages such as C++, Java and Python have, a feature that made BASIC a popular in the early ages of microcomputers. Thus the adoption of block-based languages as an introductory language is premised upon the fact that once proficient in computational thinking, learners will transition to a text based and more powerful language such as C++, Java or Python accepting some transition cost; the cost of moving from a non-text to a text-based programming world \cite{WeintropH17}. The primary goal of the Code{\em Mapper} system is to reduce this transitional cost using a novel approach.

Novice and seasoned programmers alike, often need help and they seek it from social computing platforms such as StackOverflow,  SoureForge or GitHub. The origin of the concept of crowd computing \cite{BarowyBGS17} as a distinct discipline can be traced back to its informal roots to sites such as these. While crowd computing and debugging \cite{FerranBCJ18,WatsonLG12s} opened up new ways learners and practitioners can improve and sharpen their skills, they do not directly contribute to the learners' own program development processes. In Code{\em Mapper}, we aim to leverage the crowd to aid effortless abstract program development to support the goal of reducing transitional cost from block-based to text-based coding for budding programmers.

\subsection{The Main Idea Behind {\em Code}Mapper}
\label{main}

Code{\em Mapper}'s visual interface allows learners to stay in their conceptual world similar to a block-based programming environment and yet write computer programs in a text-based imperative language such as Java, C++ or Python. In this world, solutions to problems are still a series of logical steps with well defined meanings, and any arbitrary implementation of these steps in any programming language will eventually amount to an executable program. Code{\em Mapper} consists of a database $\mathcal D$ of concepts $\mathcal C$, a concept hierarchy $\mathcal H$, a concept to program mapper $\mu$, and a program aggregator $\alpha$. A concept $c\in \mathcal C$ is an arbitrary code fragment in an imperative language in the concept hierarchy $\mathcal H$ which is organized into a generalization-specialization hierarchy DAG, e.g., merge sort and insertion sort are sorts. But merge sort itself can be broken down to the following steps, and thus is an aggregation of three concepts -- {\em divide list}, {\em merge sort a list}, and {\em merge two sorted lists}.
\begin{enumerate}
\item Divide the list in half
\item If the first half has only one element, return element, otherwise merge sort the first half
\item If the second half has only one element, return element, otherwise merge sort the second half
\item Merge both halves back together
\end{enumerate}
Each concept $c$ is either harvested from social computing sites such as StackOverflow,  SoureForge or GitHub, or are written by learners themselves, and are independent code segments. They are written in modular forms such as formal procedures or functions and not necessarily written keeping variable name or type compatibility in mind. However, they are all annotated to describe their input-output behaviors and the associated variables. Given a set of concepts, and their precedence ordering relationships, the aggregator function $\alpha$ harmonizes the variables using a schema matcher such as S-Match \cite{GiunchigliaAP12}. The code mapper function $\mu$ selects an appropriate concept while mapping a set programming steps as the merge sort above, and aims to find the most specialized concept possible before assimilating the corresponding code segment in database $\mathcal D$ to aggregate. The ultimate aggregate, an executable program $P$, is a schema heterogeneity resolved tree of concepts $T$, i.e., $P=\alpha(\mu(T))$. The role of the crowd in Code{\em Mapper}, however, is the task of placing the concepts in the hierarchy $\mathcal H$ as a community, and annotating and curating the annotations and the aggregate concepts. In that sense, Code{\em Mapper} will be supported by the developer community through crowdsourcing similar to IMPACT \cite{Mattauch2013} and Crowd Debugging \cite{ChenK2015}.

\begin{figure*}[htb!]
\centerline{\includegraphics[height=4.05in,width=.81\textwidth]{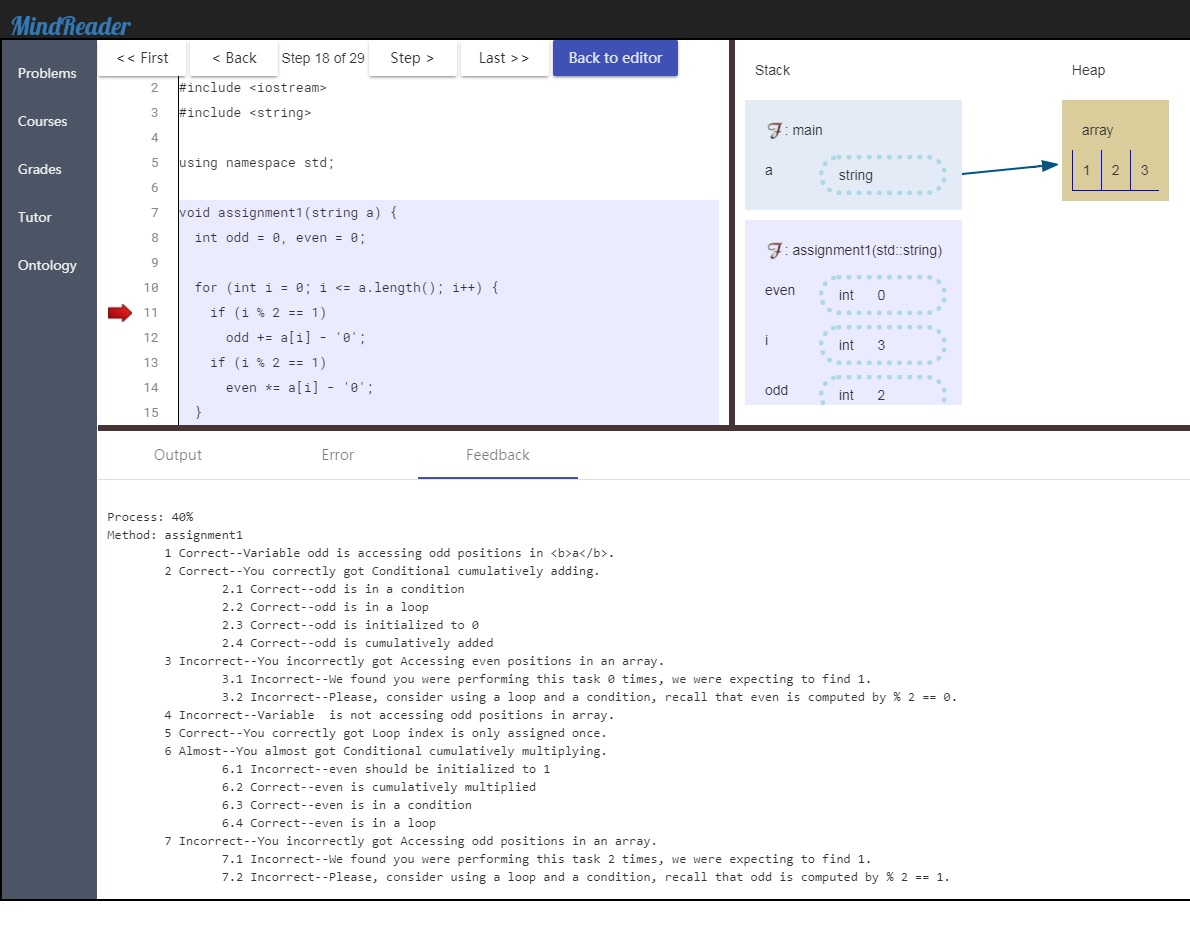}}
\caption{Mind{\em Reader} front-end.} \label{mr}
\end{figure*}

While Code{\em Mapper} may have significant similarities with Scratch and  Blockly, it is fundamentally different. In Code{\em Mapper}, students write codes or import codes in text-based languages, and thus they must be somewhat familiar with the syntax, the degree of which could be quite wide -- from elementary to well-versed. The programming itself is still visual in conceptual steps. Users of Code{\em Mapper} are able to pick concepts from the hierarchy $\mathcal H$, and sequence them to describe a solution, and then with the click of a button, the entire program is written for them. By executing, and observing the computation, they are able to refine and improve the program at the highest level of perfection possible in the database $\mathcal D$, allowed by the learner's sophistication.

\subsection{The Mind{\em Reader} System}
In recent years, numerous conversational intelligent tutoring and assessment systems (CITS) have been developed. Numerous meta-analyses of the research findings suggest that the choice of tutoring systems does not impact the learning outcome substantially and that the learners psychological make up, learning habits and anxieties often negated the technology choice \cite{MaANL14}, although the use of learning management systems (LMS) always helped. Keeping these in mind, we have developed a fully autonomous online CITS, called the Mind{\em Reader}, that includes an array of functions and features to help K-12 and early college STEM learners develop computational thinking. Our system draws upon the combined experiences of the Python Tutor \cite{Guo13} in its role as a program execution visualizer, and feedback generation techniques proposed by Martin et al. \cite{MartinPSR17} to aid comprehension through misconception elimination. Figure \ref{mr} shows the Mind{\em Reader} interface giving feedback to a student on her submitted assignment online. Mind{\em Reader} also supports and assembles an array of powerful features under one umbrella to offer a constellation of learning tools for both teachers and students. In particular it currently supports tutoring and assessment of C++, a visual programming by example language called {\em Patch} \cite{Patch-ICWL-2017}, an automated and conversational instructional agent called {\em vTutor} \cite{vTutor-ICWL-2018}, and a social peer support system called Open{\em School} \cite{OpenSchool-ICALT-2018}. The Code{\em Mapper} system introduced in this paper is designed to be an integral part of the Mind{\em Reader} system as well to help learners transition from block-based languages to C++.

\section{Background and Related Research}

Code{\em Mapper} draws upon experiences of past research in teaching computer science (CS) and web-based teaching in general, programming with the crowd, and knowledge summarization and structuring. Past research on CS education is varied but the one issue that is probably settled is that there are no simple shortcuts to CS education \cite{HassinenM2006} and the jury is still out on what is truly effective \cite{NooneM2017,HermansA17}. Apart from the content and platform debate, researchers are also experimenting with various pedagogy to make the language selection more effective using new learning environments such as immersive learning, game-based learning, blended learning, personalized, self-regulated and self-paced learning, social learning, peer learning, pair programming, etc.

While leveraging crowd \cite{BarowyBGS17} for computing is gaining steady traction, it has not been a focal point for CS education yet except for a few notable tangential research \cite{KayaO15,FerranBCJ18}. Our own recent effort in Open{\em School} \cite{OpenSchool-ICALT-2018}, and the current research on Code{\em Mapper} aim to leverage social computing platforms for the purpose of teaching and learning. Similarly, efforts in structuring knowledge to aid computer programming education is rare although there have been some effort in using ontologies or structured knowledge to aid CS education \cite{YenW17}. However, the idea of concept hierarchy that we are pursuing has not been directly researched in computing literature. In bio-informatics, functional similarity of biological concepts such as protein functions, miRNA and genes have been on going for some time. For the current edition of Code{\em Mapper}, we plan to use the crowd for such determination and annotation instead of a more algorithmic approach. In an algorithmic approach, functionally summarizing graphs appears to be a logical choice \cite{MiaoQW2017} though we have borrowed ideas from summarizing ontologies based on RDF sentence graphs \cite{ZhangCQ2007}.

\section{Code{\em Mapper}: Approach and Uniqueness}

Our main goal is to develop a programming platform that is visual, concept oriented, flexible and extensible. In this platform, concepts are searchable, using simple keyword queries, and selectable from a DAG like concept hierarchy in the Code{\em Mapper} dash board. While learners use an imperative language such as C++ or Java to program, they still use block-like icons that represent a concept and describe them. A concept can be simple or an aggregation of several concepts. Thus the expansion of a concept generally returns a code tree. Our goal is for Code{\em Mapper} to have the ability to allow novice programmers to tinker around with ideas, focus on thinking like a programmer, rather than concentrating on syntax. We achieve this by allowing users to string together concept icons in the DAG to create runnable programs in a programming language of choice. At the same time, we aim for the system to be powerful enough so that users can create complex programs in the form of a graph-like structure and generate functional programs to solve challenging problems. As mentioned earlier, the annotation of nodes in the DAG by the crowd is designed to educate Code{\em Mapper}s users about the icon functionalities and properties so that the first two goals of the system can be met accurately.

\subsection{Concept Based Programming}

\begin{figure}[h!]
\centerline{\includegraphics[height=2.025in,width=.405\textwidth]{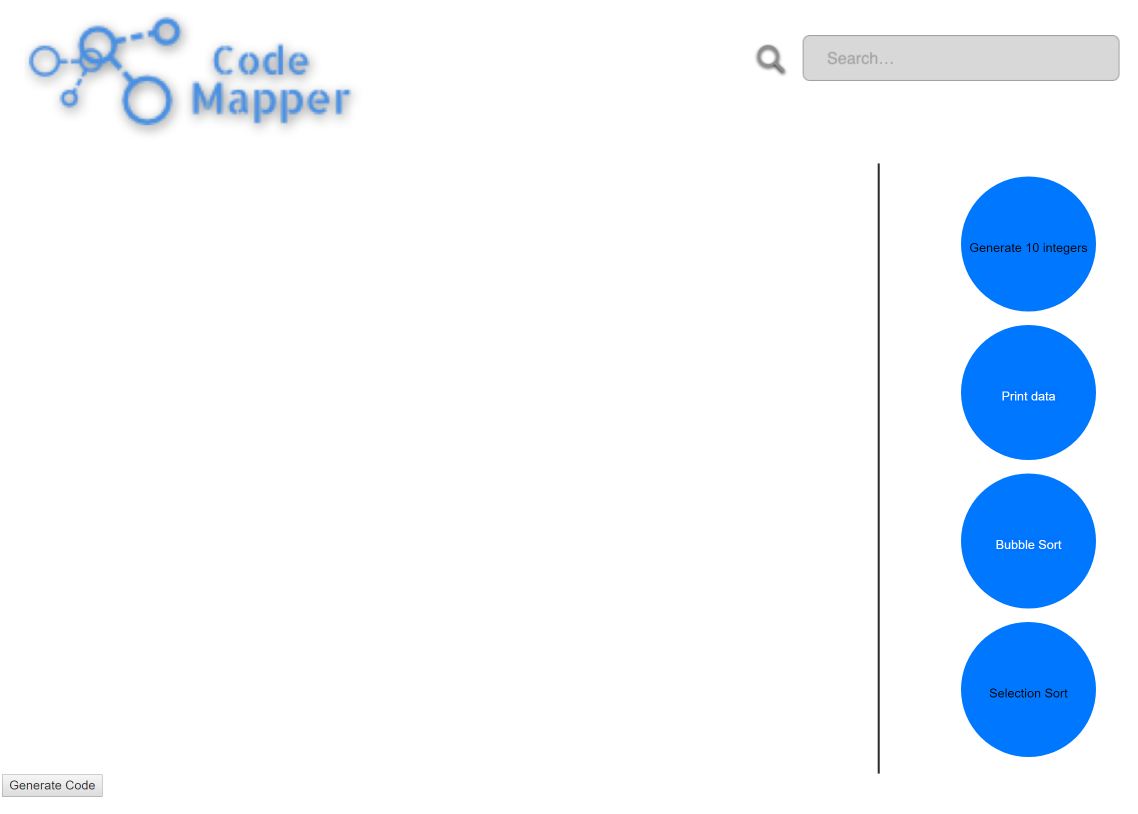}}
\caption{Code{\em Mapper} front-end.} \label{cmui}
\end{figure}

Code{\em Mapper} uses the idea of building a directed acyclic graph that directly maps to a functional program. Requiring a user program to be a DAG of icons, a concept DAG, ensures a definitive start and end point, all the while preventing a non terminating loop of tasks to occur. The front-end interface of Code{\em Mapper} is shown in figure \ref{cmui} which is the gateway to all its functional features.

\begin{figure}[h!]
\centering
\includegraphics[height=2.925in,width=.405\textwidth]{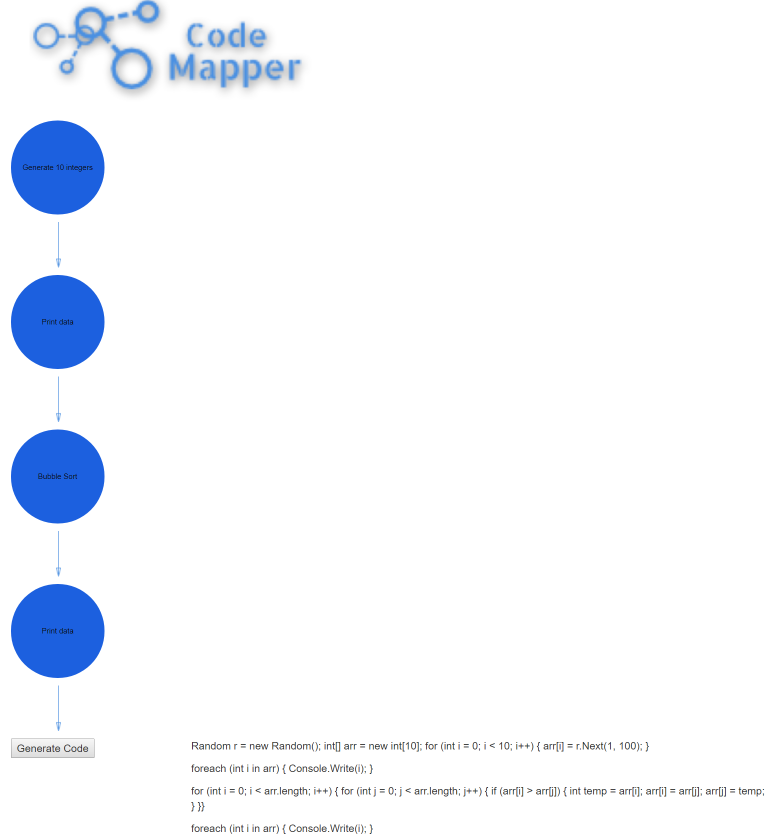}
\caption{Program induced by a concept DAG.}\label{cdag}
\end{figure}

Users write programs in Code{\em Mapper} by selecting concepts from the concept hierarchy on the dash board, dragging and dropping them on the canvas, and connecting them using arrows to create a precedence graph. They then can select a programming language for Code{\em Mapper} to map the concept graph into an executable code. In figure \ref{cdag}, a C\# mapping is shown. It is interesting to note how the flow of the node icons directly correlates to the structure of the generated program. Code{\em Mapper} being part of the Mind{\em Reader} system, users are now able to pipe this code into the IDE for compilation and execution. Alternately, they can also copy the code directly from Code{\em Mapper} and paste into applications such as Visual Studio or Unity3D editor.

\begin{figure}[!htb]
\centerline{\includegraphics[height=2.025in,width=.405\textwidth]{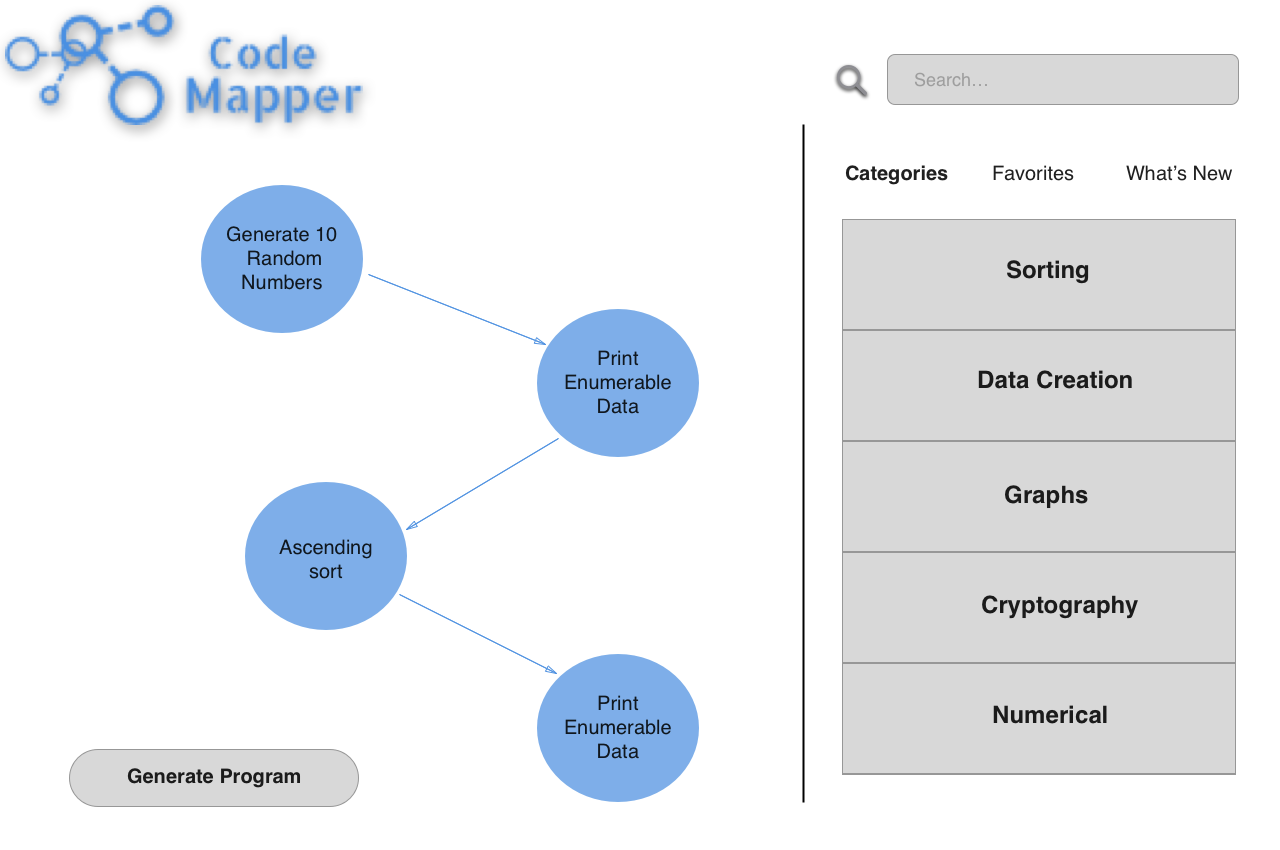}}
\caption{Concept-based sorting program implementation.} \label{finalconcept}
\end{figure}

Writing computer programs using an interface such as Code{\em Mapper} has several advantages. In such an environment, learners are able to think about the program from a high level, and focus more on the end goal while programming at an almost native language level. In this approach, users are also able to develop programs using close to everyday language rather than programming languages, so that they are more focused on problem solving as opposed to programming language idiosyncracies. In Code{\em Mapper}, the user has the opportunity to develop the concept of thinking like a programmer as they piece together these complex concept graph structures, and focus on designing a logical and most optimized solution available. Code{\em Mapper} autonomously takes care of the optimization using the recursively defined concept hierarchy by selecting the most specific concept. Figure \ref{finalconcept} shows the conceptual implementation in Code{\em Mapper} of a sorting program by a learner.

\subsection{Database Powered by Crowd-Sourcing}
\label{db-cs}

\begin{figure*}[!htb]
\centerline{\includegraphics[height=4.5in,width=.81\textwidth]{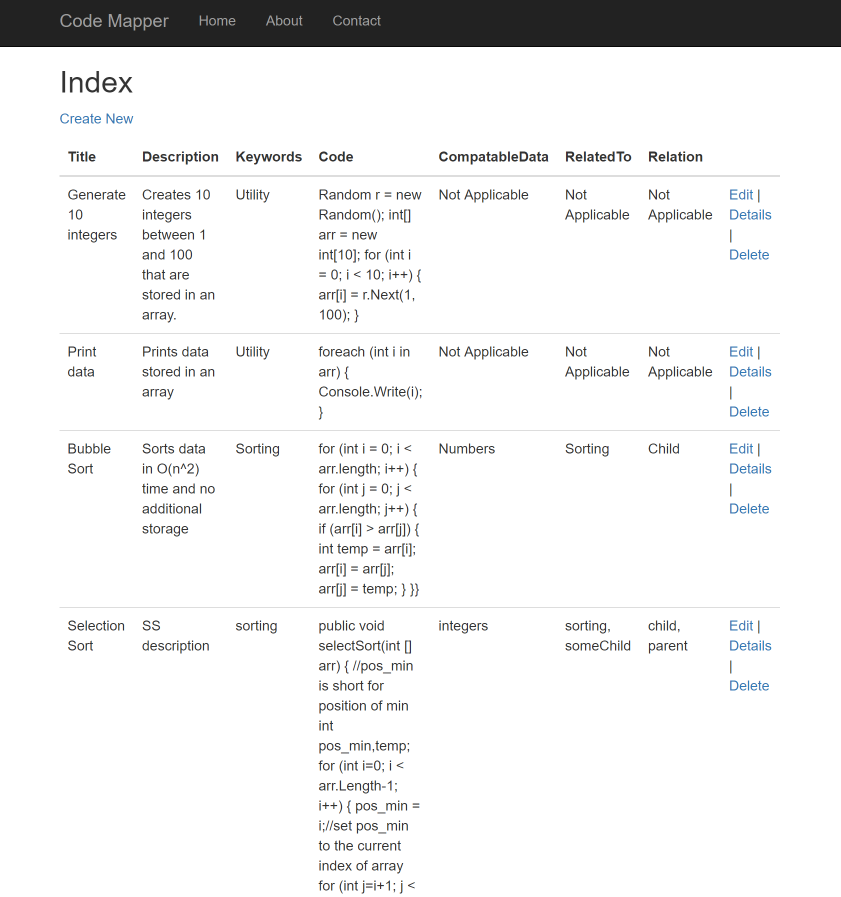}}
\caption{Detail database view of concepts.} \label{nodedata}
\end{figure*}

\begin{figure}[h!]
\centerline{\includegraphics[height=2.25in,width=.405\textwidth]{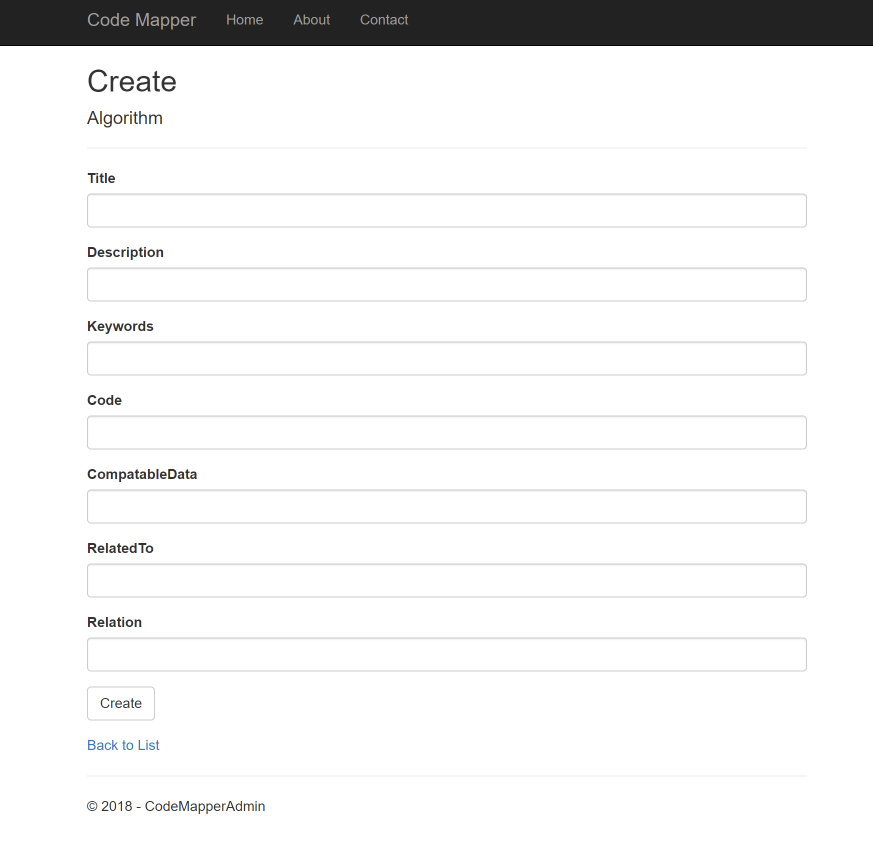}}
\caption{Data entry form for crowd annotation.}\label{entry}
\end{figure}

Every code segment associated with a terminal concept is also represented using a graph, called the program dependence graph (PDG) \cite{FerranteOW87}. Clustering of these graph structures helps organize them in the concept hierarchy into functionally similar nodes as structurally similar graphs exhibit similar execution behaviors. However, structurally dissimilar graphs can also be functionally identical. For example, though the PDGs corresponding to insertion sort and quick sort are structurally different, they are functionally identical -- sorting. It therefore becomes necessary that we engage the crowd to annotate each concept node, and manually establish their functional similarity. Figure \ref{nodedata} shows the annotated information available for each of the concepts entered using the form in figure \ref{entry}. Furthermore, the granularity of concepts could also be varied. For example, it need not be a complete procedure. We could also have a concept called {\em counter loop}, specialization of which could be a {\em for counter loop} and a {\em while counter loop} with respective {\em for} and {\em while} statement incarnations.

\begin{figure}[!htb]
\centerline{\includegraphics[height=1.8in,width=.405\textwidth]{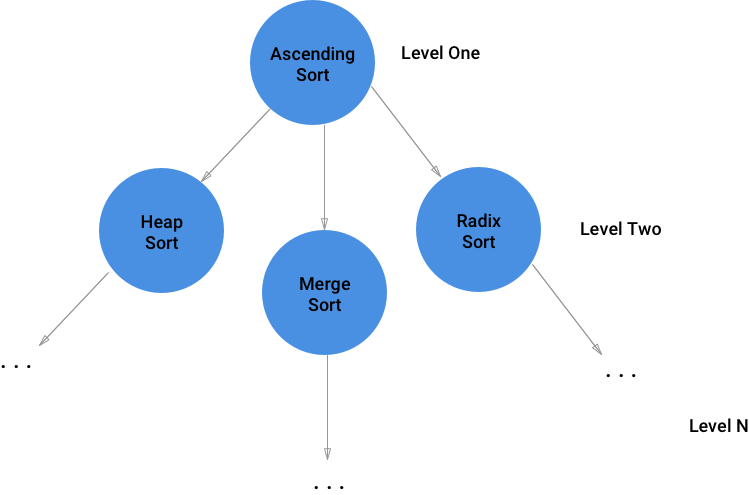}}
\caption{Example concept hierarchy.} \label{graph}
\end{figure}

Manual curation by crowd is also necessary since concepts can be aggregations of simple concepts (called terminal), or complex concepts (concepts defined using other simple or complex concepts recursively). Most complex concepts that represent similar functions cannot be clustered using PDG clustering. For example, the node structure in figure \ref{graph} is created or curated by the crowd to show heap, merge and radix sorts each to be a special type of ascending sort. As discussed in section \ref{main}, merge sort is a complex concept, and so is quick sort. While it is conceivable that most merge sort and quick sort PDGs are group wise similar, it is unlikely for the merge sort and quick sort PDGs to be similar, although they function similarly. The annotations and their position in the concept hierarchy thus becomes important in Code{\em Mapper}, allowing the system to use the specificity and referencing (aggregations) relationships among concepts to make smart decisions in synthesizing high quality programs. From this perspective, concepts and their realization in Code{\em Mapper} have strong similarity with those of knowledge representation of ontologies, and we envision Code{\em Mapper} to be similar to sites such as StackOverflow or Github, in which crowd sourcing fuels it's success.

\section{Open Source Implementation}

We have developed Code{\em Mapper} as an open source system so that it can be developed as a community project and leverage the combined ideas of all. The source code of Code{\em Mapper} project publicly available at https://github.com/maholeycowdevelopment/CodeMapper. It is therefore open to contributions and will soon be under the MIT license. We plan to create a suggestions page to solicit ideas about desired features and implementation strategies to facilitate its continuous development.

\subsection{Software and Platform Tools}

Code{\em Mapper} is built with Microsoft's ASP.NET Core 2.0 platform with Facebook's front-end library \mbox{React}. This choice was motivated by its flexibility and extensibility. This platform allows flexible addition and removal of functionality, and thus supports incremental design while focusing on the end goal. Facebook's React was motivated by the need for dynamic user experience and user interface operations, and for leveraging React's ability to allow building reusable components. Several command line tools such as node package manager, .NET common line tools, and yarn package manager were also used. For back-end system development, we have used either a yarn or npm development server, and the IIS Express along with the .NET command line servers.

\subsection{Hardware and Operating System}

Code{\em Mapper} was implemented on MacBook Pro running macOS High Sierra version 10.13.1, having a 2.8 Ghz Intel Core i7 processor, 16 GB 2133 MHz LPDDR3, Radeon Pro 555 2 GB Intel HD Graphics 630 1536 MB. Much of the development was done on a virtual machine utilizing Parallels software. Code{\em Mapper} is compliant with most widely used browsers such as Google Chrome, Microsoft Edge, Internet Explorer, Safari, and Fire Fox. We tested the software on different machines of varying configurations and did not experience any performance related glitches to date.

\subsection{Online Code Harvesting}

\begin{figure}[!htb]
\centerline{\includegraphics[height=2.25in,width=.405\textwidth]{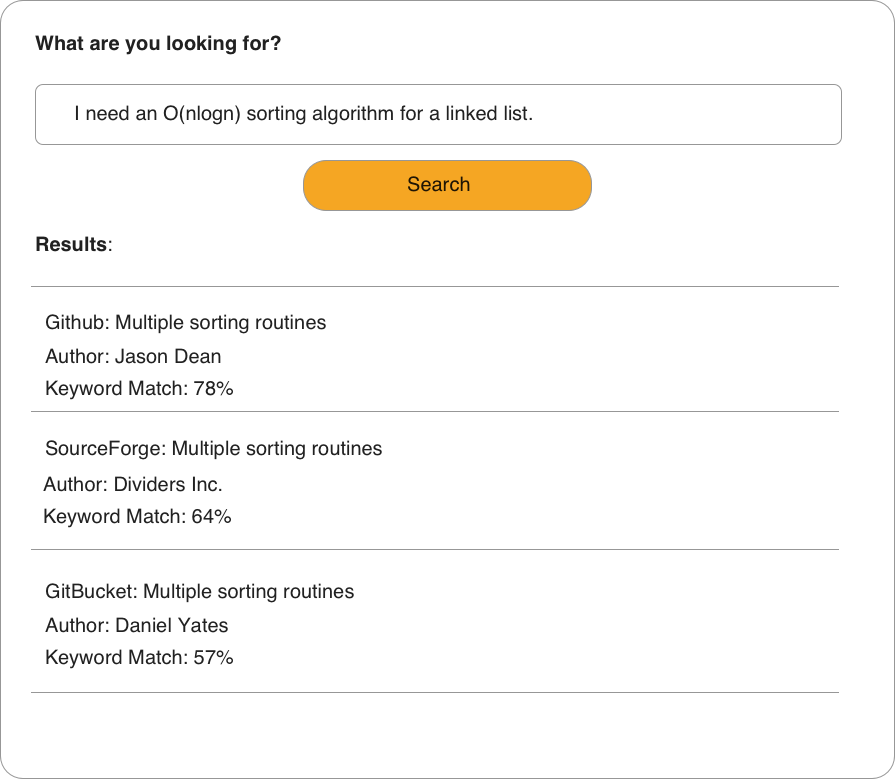}}
\caption{Online code search and find interface.}
\label{code-search}
\end{figure}

In order to make Code{\em Mapper} a truly powerful tool, users should be able to scrape the internet for code snippets to be used in node creation. It supports the search and find interface shown in figure \ref{code-search} for this purpose. In this interface, search for codes can be initiated using a brief description of target code. Code{\em Mapper} then initiates search using the APIs provided by sites such as GitHub, SourceForge, GitBucket, etc. The API's are then interrogated using NLP techniques to pick out specific terms. Once the results are returned in a list as shown, users pick and choose snippets of code as needed to create nodes. Users are able to copy the code snippets into the system database for cataloging in ways discussed in section \ref{db-cs}.

\subsection{Experimental Evaluations}

We have tested Code{\em Mapper} positively for its functional correctness. In our lab test,  programming.  In our user experience tests, learners without significant coding experience found it useful and easier to understand. The most significant insight from this process was that students showed improvement in thinking about program design without the nuances of a language's syntax. We
also noted that the majority of students stated they were able to solidify programming concepts and develop a deeper understanding of programming by looking at the produced code and the DAG they built. However, a more systematic evaluation is being planned.

Contingent upon its adoption, it would be a significant step for the viability and success of crowd-sourcing based rapid prototyping of programs, and a program development platform for novice and veteran programmers alike. We also envision its adoption as an intelligent and shared crowd enabled community software development platform where users can grow their skills and mature together. The success of its graph summarization could also be useful in the area of community debugging research.

\subsection{Empowering Novice Programmers}

Since Code{\em Mapper} focuses on incremental code development using concepts, in ways similar to blocks in Scratch and Blockly, learners are able to focus more on what to solve, and not how to solve. From this viewpoint, Code{\em Mapper} is closer to the declarative programming paradigm. Learners have the opportunity to build and reuse concepts, and their implementations, themselves or leverage the crowd. Since the actual integration of the independently constructed code segments and resolution of their disparities are performed by Code{\em Mapper} using well established techniques, novice programmers need not invest too much time to construct new programs the way traditional programmers still do. The saved time can now be used to explore alternate implementations, or discover the pitfalls of the design at hand because a low quality (performance or construction wise) concept was chosen. Once a concept graph is built, the mapping to real code by Code{\em Mapper} and the exploration of the generated codes will  allow users to make connections between syntax and semantics.

Furthermore, since Code{\em Mapper} is powered by crowd sourcing, learners can observe the emergence of new concepts and their evolution that is not possible in static and non-evolving system such as Blockly or Scratch. The modularity inherent in the concepts and their implementation allows research and comprehension of algorithms in a structured fashion once they are expanded and helps learners sharpen their abilities as a programmer. Learners also have the opportunity to take part in developing some of the concepts and their annotations introducing them to the process of open source system development. Such collaborative experience in system building will be beneficial for learners' abilities to contribute to school projects, internships, personal projects and so on.

\subsection{Industrial Relevance}

We believe that industrial developers can also utilize Code{\em Mapper} for serious system building. In many instances, engineers are looking for strategies to rapidly prototype concepts that they want to implement in an application, but do not want to spend the man hours needed to research how to do this. They can now turn to Code{\em Mapper} and create a quick prototype of the desired features, test it, and then make modifications as necessary. This will benefit companies, users, and the technology community. As the community database of Code{\em Mapper} grows, it is not far-fetched to imagine Code{\em Mapper} becoming a vehicle for professional system development in the foreseeable future.

\subsection{Program Dependence Graph Summarization}

One of the components of Code{\em Mapper} is the concept hierarchy $\mathcal H$ that organizes all the concepts learners use to develop conceptual solutions to problems. The accuracy of the hierarchy is important for two main reasons. First, the hierarchy is used by the mapping function $\mu$ in order to select most up to date and appropriate code segments. Second, structuring the concepts in a specialization-generalization hierarchy helps learners select the most abstract (general) concept possible, which will ultimately map to the most specific code anyway. This independence allows learners to think solutions and not techniques. As mentioned, while technologies exist to cluster similarly functioning codes that have similar PDGs, no technology is suitable to cluster similarly functioning codes with divergent PDGs. We are thus unavoidably dependent on the power of the crowd for subjective and manual clustering toward the construction of the concept hierarchy.

To automate the process, we believe a deductive knowledge base can be created and curated by the crowd, the rules of which can then be used to identify similarly functioning divergent codes (thus vastly different PDGs) and clustered as similar concepts. In fact, it is possible to express the definition of a merge sort, for example, as described in section \ref{main} as a set of logic rules. Then these rules can be mapped to codes that can faithfully implement those rules, not necessarily the same way. An experimental implementation has been discussed and utilized in Mind{\em Reader} \cite{MindReader-ICALT-2017}, and is currently undergoing evaluation. But this experimental system is manual and must be orchestrated by users and thus, is not scalable. Our hope is that leveraging the power of crowd, it will be possible to scale the approach and simplify the concept hierarchy generation for Code{\em Mapper}.

\subsection{Limitations and Future Improvements}
\label{disc}

Being crowd reliant, Code{\em Mapper} inherits typical drawbacks of this paradigm, yet being a community system, it also has some serious responsibility. For example, the annotation accuracy is largely curator dependent and a weak, incompetent or malicious curator may jeopardize the system, and thus a users ability to construct meaningful and useful programs. One of two possible solutions appears appropriate. The standard approach practiced in similar circumstances such as GitHub is to accept only administrator curated annotations, and allow local overriding by any user on their own without affecting the community annotation. The second approach is to use an inaccurate model where the reliability of an annotation is dependent upon the combined credibility of all the curators who contributed to the annotation, an approach adopted in CrowdCure system \cite{JamilS-CrowdCure-DAPD-2018}.

The current edition of Code{\em Mapper} does not allow hypothetical concepts, or their implementation, to be included in a concept graph for a solution. It also does not allow mixing implementations in mixed languages. But it may so happen that a learner is only familiar with Scratch and is able to implement a concept using it well, but not so much using C++. Allowing her to add a Scratch implementation, or choosing an available Python implement, to other components of her concept graph in C++, could offer a workable solution. For this to be executable, a translation from Scratch or Python to C++, or vice versa, will be necessary. A system such as Flowgorithm \cite{Flowgorithm} actually allows cross language mapping from abstract flow charts, and a similar approach can be used in Code{\em Mapper} as well.

The current implementation also does not support partial evaluation of concepts within a larger concept graph to test ideas locally. This also means that the solution must be fully conceived and tested, and no part of it can be independently evaluated. We would like to device a mechanism in the future to support such incremental evaluation and idea exploration that we believe will be beneficial to novice and seasoned programmers alike. In that vein, we also believe that an auto suggestion of possible concepts can prove to be useful. In particular, given a problem, certain concepts being considered by the learner may not make sense, and rather than waiting for Mind{\em Reader} to catch the discrepancy and give feedback later, Code{\em Mapper} may suggest making alternative choice and avert an eventual failure. In this role, Code{\em Mapper} will essentially be acting as an online tutor for novice programmers.

\section{Conclusion and Future Research}

The focus of this paper has been to articulate a novel idea of softening the transition cost of novice programmers and \mbox{K-12} STEM learners from block-based languages to text-based imperative languages and foster their computational thinking. While the Code{\em Mapper} system is at the prototype stage, most of its core components have been developed and tested, but its integration with the sister system Mind{\em Reader} remains as a future research. We have also discussed some of the planned future improvements in section \ref{disc} and its known limitations. We believe that though Code{\em Mapper} is a part of the overall Mind{\em Reader} system, it stands on its own as an intelligent web application that has significant potential to help impart quality computational thinking education.

%\newpage

\bibliographystyle{abbrv}
%\bibliography{/users/jamil/dropbox/bib-db/bib-db-general,/users/jamil/dropbox/bib-db/our-publications}
%\bibliography{/users/hasan/dropbox/bib-db/bib-db-general,/users/hasan/dropbox/bib-db/our-publications}

\begin{thebibliography}{10}

\bibitem{BarowyBGS17}
D.~W. Barowy, E.~D. Berger, D.~G. Goldstein, and S.~Suri.
\newblock Voxpl: Programming with the wisdom of the crowd.
\newblock In {\em Proceedings of the 2017 {CHI} Conference on Human Factors in
  Computing Systems, Denver, CO, USA, May 06-11, 2017.}, pages 2347--2358,
  2017.

\bibitem{CCSS}
CCSS.
\newblock Common core state standards initiative.
\newblock \url{http://www.corestandards.org/}, 2010.
\newblock Accessed on April 19, 2017.

\bibitem{ChenK2015}
F.~Chen and S.~Kim.
\newblock Crowd debugging.
\newblock In {\em Proceedings of the 2015 10th Joint Meeting on Foundations of
  Software Engineering}, ESEC/FSE 2015, pages 320--332, New York, NY, USA,
  2015. ACM.

\bibitem{Flowgorithm}
D.~Cook.
\newblock Flowgorithm home page.
\newblock \url{http://www.flowgorithm.org/}.
\newblock Accessed: June 17, 2018.

\bibitem{FerranBCJ18}
S.~Ferr{\'{a}}n, A.~Beghelli, G.~H. C{\'{a}}nepa, and F.~Jensen.
\newblock Correctness assessment of a crowdcoding project in a computer
  programming introductory course.
\newblock {\em Comp. Applic. in Engineering Education}, 26(1):162--170, 2018.

\bibitem{FerranteOW87}
J.~Ferrante, K.~J. Ottenstein, and J.~D. Warren.
\newblock The program dependence graph and its use in optimization.
\newblock {\em {ACM} Trans. Program. Lang. Syst.}, 9(3):319--349, 1987.

\bibitem{GiunchigliaAP12}
F.~Giunchiglia, A.~Autayeu, and J.~Pane.
\newblock S-match: An open source framework for matching lightweight
  ontologies.
\newblock {\em Semantic Web}, 3(3):307--317, 2012.

\bibitem{Guo13}
P.~J. Guo.
\newblock Online python tutor: embeddable web-based program visualization for
  cs education.
\newblock In {\em The 44th {ACM} Technical Symposium on Computer Science
  Education, {SIGCSE} '13, Denver, CO, USA, March 6-9, 2013}, pages 579--584,
  2013.

\bibitem{HassinenM2006}
M.~Hassinen and H.~M\"{a}yr\"{a}.
\newblock Learning programming by programming: A case study.
\newblock In {\em Proceedings of the 6th Baltic Sea Conference on Computing
  Education Research: Koli Calling 2006}, Baltic Sea '06, pages 117--119, New
  York, NY, USA, 2006. ACM.

\bibitem{HermansA17}
F.~Hermans and E.~Aivaloglou.
\newblock To scratch or not to scratch?: {A} controlled experiment comparing
  plugged first and unplugged first programming lessons.
\newblock In {\em Proceedings of the 12th Workshop on Primary and Secondary
  Computing Education, WiPSCE 2017, Nijmegen, The Netherlands, November 08 -
  10, 2017}, pages 49--56, 2017.

\bibitem{MindReader-ICALT-2017}
H.~M. Jamil.
\newblock Automated personalized assessment of computational thinking {MOOC}
  assignments.
\newblock In {\em Proceedings of The 17th IEEE International Conference on
  Advanced Learning Technologies, {ICALT} 2017, Timisoara, Romania, July 3-7},
  pages 261--263, 2017.

\bibitem{Patch-ICWL-2017}
H.~M. Jamil.
\newblock Visual computational thinking using {{\em Patch}}.
\newblock In {\em Proceedings of The 16th International Conference on Web-based
  Learning, {ICWL} 2017, Cape Town, South Africa, September 20-22}, pages
  208--214. Springer, LNCS 10473, 2017.

\bibitem{OpenSchool-ICALT-2018}
H.~M. Jamil.
\newblock A free-choice social learning network for computational thinking.
\newblock In {\em Proceedings of The 18th IEEE International Conference on
  Advanced Learning Technologies, {ICALT} 2018, Mumbai, India, July 9-13},
  2018.
\newblock To appear.

\bibitem{vTutor-ICWL-2018}
H.~M. Jamil, X.~Mou, R.~B. Heckendorn, C.~L. Jeffery, F.~T. Sheldon, C.~S.
  Hall, and N.~M. Peterson.
\newblock Authoring adaptive digital computational thinking lessons using
  {vTutor} for web-based learning.
\newblock In {\em Proceedings of The 16th International Conference on Web-based
  Learning, {ICWL} 2018, Chiang Mai, Thailand, August 22-24}. Springer, 2018.
\newblock To appear.

\bibitem{JamilS-CrowdCure-DAPD-2018}
H.~M. Jamil and F.~Sadri.
\newblock Crowd enabled curation and querying of large and noisy text mined
  protein interaction data.
\newblock {\em Distributed and Parallel Databases}, 36(1):9--45, 2018.

\bibitem{KayaO15}
M.~Kaya and S.~A. {\"{O}}zel.
\newblock Integrating an online compiler and a plagiarism detection tool into
  the moodle distance education system for easy assessment of programming
  assignments.
\newblock {\em Comp. Applic. in Engineering Education}, 23(3):363--373, 2015.

\bibitem{KollingBA15}
M.~K{\"{o}}lling, N.~C.~C. Brown, and A.~AlTadmri.
\newblock Frame-based editing: Easing the transition from blocks to text-based
  programming.
\newblock In {\em WiPSCE}, pages 29--38. {ACM}, 2015.

\bibitem{KunkleA16}
W.~M. Kunkle and R.~B. Allen.
\newblock The impact of different teaching approaches and languages on student
  learning of introductory programming concepts.
\newblock {\em {TOCE}}, 16(1):3:1--3:26, 2016.

\bibitem{MaANL14}
W.~Ma, O.~O. Adesope, Nesbit, J.~C., and Q.~Liu.
\newblock Intelligent tutoring systems and learning outcomes: A meta-analysis.
\newblock {\em Journal of Educational Psychology}, 106(4):901--918, 2014.

\bibitem{MartinPSR17}
V.~J. Martin, T.~Pereira, S.~Sridharan, and C.~R. Rivero.
\newblock Automated personalized feedback in introductory java programming
  moocs.
\newblock In {\em Proceedings of the 33rd IEEE International Conference on Data
  Engineering, San Diego, California, USA, April 19-22}, pages 1259--1270,
  2017.

\bibitem{Mattauch2013}
T.~Mattauch.
\newblock Innovate through crowd sourcing.
\newblock In {\em Proceedings of the 41st Annual ACM SIGUCCS Conference on User
  Services}, SIGUCCS '13, pages 39--42, New York, NY, USA, 2013. ACM.

\bibitem{MiaoQW2017}
Y.~Miao, J.~Qin, and W.~Wang.
\newblock Graph summarization for entity relatedness visualization.
\newblock In {\em Proceedings of the 40th International ACM SIGIR Conference on
  Research and Development in Information Retrieval}, SIGIR '17, pages
  1161--1164, New York, NY, USA, 2017. ACM.

\bibitem{NGSS}
NGSS.
\newblock Next generation science standard.
\newblock \url{http://www.nextgenscience.org/}, 2013.
\newblock Accessed on April 19, 2017.

\bibitem{NooneM2017}
M.~Noone and A.~Mooney.
\newblock First programming language: Visual or textual?
\newblock {\em CoRR}, abs/1710.11557, 2017.

\bibitem{PorterGMS13}
L.~Porter, M.~Guzdial, C.~McDowell, and B.~Simon.
\newblock Success in introductory programming: what works?
\newblock {\em Commun. {ACM}}, 56(8):34--36, 2013.

\bibitem{VihavainenAW14}
A.~Vihavainen, J.~Airaksinen, and C.~Watson.
\newblock A systematic review of approaches for teaching introductory
  programming and their influence on success.
\newblock In {\em {ICER}}, pages 19--26. {ACM}, 2014.

\bibitem{WatsonL14}
C.~Watson and F.~W.~B. Li.
\newblock Failure rates in introductory programming revisited.
\newblock In {\em ITiCSE}, pages 39--44. {ACM}, 2014.

\bibitem{WatsonLG12s}
C.~Watson, F.~W.~B. Li, and J.~L. Godwin.
\newblock {BlueFix}: Using crowd-sourced feedback to support programming
  students in error diagnosis and repair.
\newblock In {\em {ICWL}}, volume 7558 of {\em LNCS}, pages 228--239, 2012.

\bibitem{WeintropH17}
D.~Weintrop and N.~R. Holbert.
\newblock From blocks to text and back: Programming patterns in a dual-modality
  environment.
\newblock In {\em {SIGCSE}}, pages 633--638. {ACM}, 2017.

\bibitem{WingJ2016}
J.~M. Wing.
\newblock {\em Computational thinking, 10 years later}.
\newblock https://tinyurl.com/yapf5zas, Mar. 2016.
\newblock Accessed: August 31, 2017.

\bibitem{YenW17}
C.~Yen and T.~Wang.
\newblock Using self-explanation and ontology for providing proper feedbacks in
  a programming environment.
\newblock In {\em 6th {IIAI} International Congress on Advanced Applied
  Informatics, {IIAI-AAI} 2017, Hamamatsu, Japan, July 9-13, 2017}, pages
  585--590, 2017.

\bibitem{ZhangCQ2007}
X.~Zhang, G.~Cheng, and Y.~Qu.
\newblock Ontology summarization based on rdf sentence graph.
\newblock In {\em Proceedings of the 16th International Conference on World
  Wide Web}, WWW '07, pages 707--716, New York, NY, USA, 2007. ACM.

\end{thebibliography}

\end{document}